# High Performance GNR Power Gating for Low-Voltage CMOS Circuits

Hader E. El-hmaily, Rabab Ezz-Eldin*, A. I. A. Galal** and Hesham F.A.Hamed**
Electrical Engineering Dept., Badr University, Cairo, Egypt.
Electrical Engineering Dept., Beni-Suef University, Beni-Suef, Egypt*
Electrical Engineering Dept., Minia University, El-Minia, Egypt**
E-mails: hader.essam@buc.edu.eg, rabab.ezz@eng.bsu.edu.eg, Galal@mu.edu.eg, hfah66@yahoo.com

***Abstract*- A robust power gating design using Graphene Nano-Ribbon Field-Effect Transistors (*GNRFET*) is proposed using 16nm technology. The Power Gating (*PG*) structure is composed of *GNRFET* as a power switch and *MOS* power gated module. The proposed structure resolves the main drawbacks of the traditional *PG* design from the point of view increasing the propagation delay and wake-up time in low–voltage regions. *GNRFET/MOSFET* Conjunction (*GMC*) is employed to build various structures of *PG*; *GMCPG-SS* and *GMCPG-NS*. In addition to exploiting it to build two multi-mode *PG* structures. Circuit analysis for *CMOS* power gated logic modules (ISCAS85 benchmark) of 16nm technology is used to evaluate the performance of the proposed *GNR* power switch is compared to the traditional *MOS* one. Leakage power, wake-up time and power delay product are used as performance circuit parameters for the evaluation. *GMCPG-SS* performance results reveal a reduction in leakage power, delay time, and wake-up time, on average up to 88%, 44%, and 24%, respectively, and *GMCPG-NS* structure reduces the leakage power in between 69% and 92%, and wake-up time by 27-46% for different ISCAS85 power gated modules compared to the *MOSPG* structure. Both multimode PG structures are able to reduce the leakage power as compared to the other PG structures with improvement in the wake-up time by 99%.**

## I. INTRODUCTION

Nowadays, scaling down *MOSFET* to advanced technologies has a lot of difficulties such as keeping up Moore's Law due to the various challenges imposed by the extremely small feature sizes, including increased wire resistivity, significant mobility degradation, and large dopant fluctuations [1]. To sustain the high performance of the circuit design and overcome the scaling silicon channel challenges, the recent researches tend to replace a conventional channel material for transistors with new nano-materials that have extremely vital physical and electrical properties compared to Silicon, such carbon nano-tube, Graphene and Graphene nano-ribbon, to operate at low voltage with a low sub-threshold slope.

Graphene is a single atomic layer of graphite with two-dimensional honeycomb crystal lattice [2]. The lack of bandgap for two-dimensional Graphene is the main cause to limit using the large-area Graphene for integrated circuits [3]. Narrow Graphene Nanoribbon (*GNR*) stripes are adopted to increase the energy bandgap because of its symmetrical band structure, light effective mass, and direct band gap to favor tunneling [4]. Therefore, Graphene Nano-Ribbon Field-Effect Transistors (*GNRFET*) are exploited as a powerful transistor in digital logic applications instead of *CMOS* counterpart especially the low power application [5].

Power Gating (*PG*) is one of the popular techniques to save the leakage power during the low power mode (Sleep mode) to increase the speed of integrated circuits. The previous power gating techniques depend on exploiting *MOS* power switch (*MOSPS*) as a header switch and/or a footer switch to isolate the virtual supply nodes from the actual supply lines to turn off part of the power gated modules and enter Sleep mode. A number of techniques were proposed for reducing leakage power dissipation such as two-pass *PG* [6], and zigzag *PG* [7]. On the other hand, a reduction in wake-up time by using multi-mode *PG* compensates small increase of leakage power dissipation are presented in [8], [9], [10]. In the previously mentioned approaches, *low-*$V_{th}$ transistors are used inside the power gated modules to ensure higher performance during the Active mode of the circuit. *High-*$V_{th}$ sleep transistors are located between the power gated modules and the power supplies to reduce leakage power during the Sleep mode. The performance of the traditional power gating is degraded in the low-voltage region as a result of *high-*$V_{th}$ of the sleep transistors which leads to reduce the operation frequency and rise wake-up time rapidly at the low voltage [11].

In this paper, *MOSPS* in the traditional *PG* structure is replaced by *GNR* Power Switch (*GNRPS*) as a footer to switch the power supply of power gated module as shown in Figure 1 to improve the performance of the circuit with the advanced technologies. The evolved *PG* based on *GNRFET/MOSFET* Conjunction (*GMC*) is a robust design that resolves with the previous problems and minimizes the area overhead. Using *GMC* to build *PG* (*GMCPG)* and compare the performance of the proposed design to the traditional *PG* using different *PG* techniques as shown in the following sections.

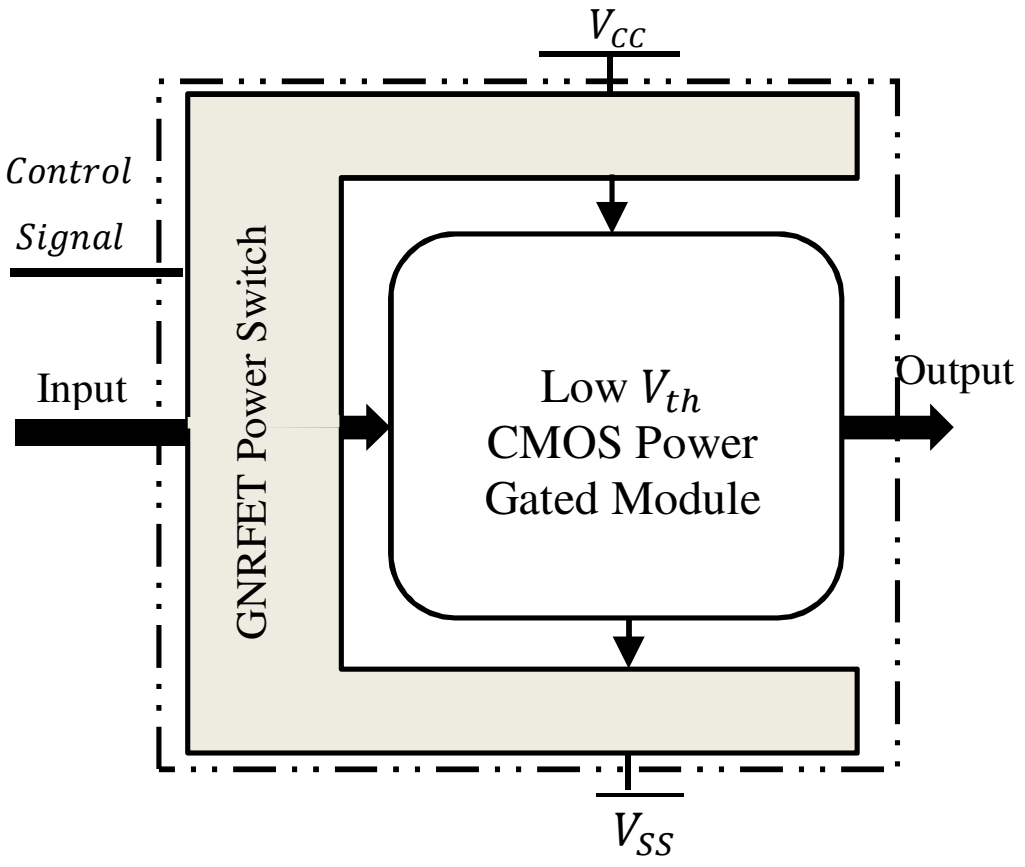

Figure 1. Power Gating Diagram using *GNR* Power Switch

The paper is organized as follows. In Section **II**, the model of *GNRFET* is adopted. Different structures for power gating based *GMC* are described in Section **III**. Two multimode power gating designs based on *GMC* are presented in Section **IV**. In Section **V**, simulation results are provided. Conclusions are demonstrated in Section **VI**.

## II. *GNRFET* Model

Graphene is used as a channel material for high speed *FET* instead of *MOSFET*. Graphene outperforms Silicon from the point of view carrier mobility, carrier concentration and thermal conductivity [12]. In addition, Graphene has automatically thin planar structure than Silicon to increase ON-current (due to a larger transconductance) and a maximize $I_{ON}/I_{OFF}$ current ratio [12]. On the other hand, the lack of bandgap called "dispersionless band" is the main cause to limit using the Graphene Field Effect Transistor (*GFET*) for integrated circuits. Mainly because the conduction and valence bands impinged each other which make the band of Graphene structure like a cone-shaped, as shown in Figure 2(a). The cone shaped leads to a zero bandgap or negative transconductance, which is undesirable property. Consequently, *GFET* is not suitable as logic transistors since the switch efficiency is very low, where the switch efficiency is defined as the ratio of drain current in the *ON* state ($I_{ON}$) to the current at *OFF* state ($I_{OFF}$). In addition, there is another problem from using *GFET* in digital application beside the bandgap issue that *GFET* requires a relatively high voltage to switch the transistor ON as a result of using a significantly thick oxide in fabrication [13].

One of the most interesting techniques to opening a bandgap is forming Graphene into narrow ribbons as an alternative channel material in FET [12] [14] [ 15]. As shown in Figure 2(b), Fermi level ($E_{F2}$) is located in the center of the energy bandgap and the total number of electrons in conduction band and the holes in valence band are minimized which leads to open bandgap and reduce the leakage current. Decreasing the gate voltage shifts the Fermi level into the gap, the carrier density in the channel decreases continuously and rapidly, and the transistor switches OFF. Creating Graphene nano-ribbon power switch would need to behave as a semiconductor, which is the key to the *ON–OFF* switching operations performed by electronic components.

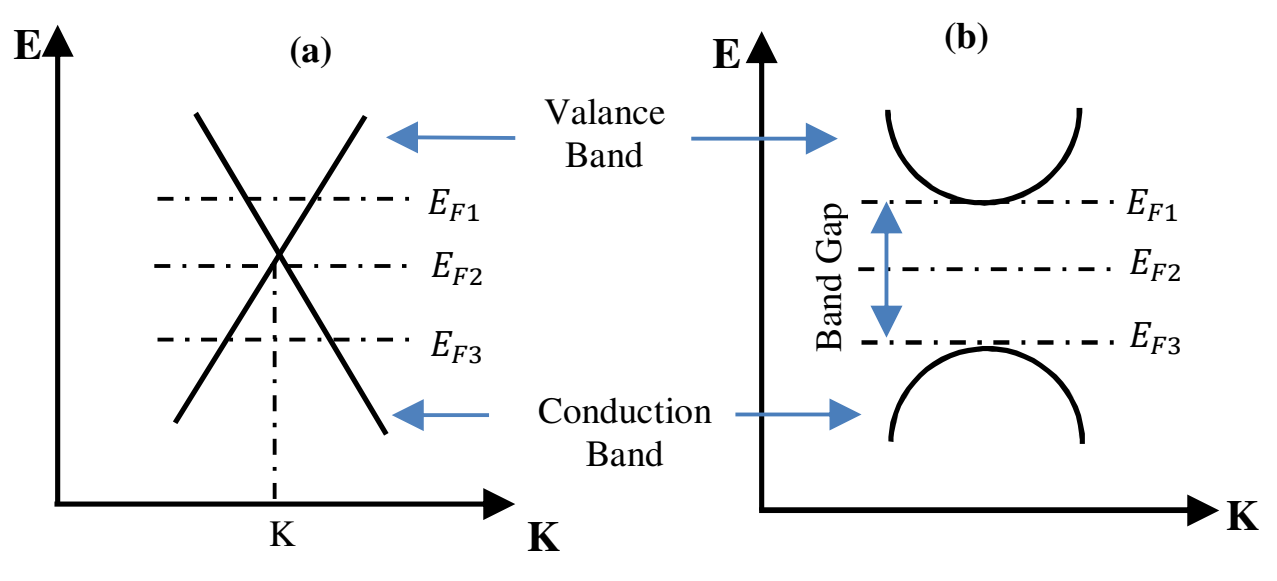

Figure 2. (a) Schematic band diagram of lack of-band-gap large-area graphene. (b) Band diagram of a semiconductor.

Semiconductors for *GNRFET* are defined by their bandgap, if there is a small band gap, then the flow of electrons from valence to conduction band is possible only if an external energy is supplied [17]. When the bandgap is wide enough, the flow of electrons from valence band to conduction band becomes few. Consequently, the bandgap needs to be large enough to allow reducing the leakage current when the *GNRFET* transistor in OFF state.

From the other point of view, the nano-ribbon chemical structure and the edge structure are extremely important to define the electrical properties of *GNRFET*. *GNR* are combination of metallic and semiconducting materials. Armchair and zigzag are the main structures of *GNR*. Zigzag *GNRs* are always metallic while armchair *GNRs* can be either a conductor or a semiconductor depends on the type of the edge boundary along *GNR*. The edge boundary represents by a chiral vector $(N,0)$ where $N$ is the number of dimer lines for the armchair. Therefore, *GNRs* are semiconductor with wide energy bandgap and highest $I_{ON}/I_{OFF}$ ratio when $N = 3P + 1$ such as $N = 3P$ such as $N = 6, 9, 12, 15,$ and 18 and $N = 10, 13, 16,$ and 19 where $P$ is an arbitrary integer [5]. While *GNRs* behave as a conductor with zero energy bandgap when $N = 3p + 2$ such as $N = 8, 11, 14,$ and 17 [5]. Armchair chirality *GNR* is used as a channel material to increase the drive strength and to form conducting contacts. *GNR* is characterized by the low sub-threshold swing, high switch efficiency, high carrier mobility [15].

Gate, Source, Drain and Substrate are the four terminals for *GNRFET*. The channel under the Gate terminal is injected an un-doped *GNR* while the reservoirs between the Gate and the wide contact are injected the heavily doped GNR with a doping fraction ($f_{dop}$) to give the Gate terminal more control over the channel region [5]. Doping the reservoirs with donors produces N-*GNRFET* akin to *NMOS* because the current is dominated by electron conduction. While, doping the reservoirs with acceptors leads to the current of P-*GRNFET* to be dominated by hole conduction and resemble a *PMOS*.

The equivalent circuit model of *GNRFET* is shown in Figure 3. The model consists of one current source $I_{DS}$ which is exploited for the DC behavior when the *GNR* channel charges or discharges due to a voltage-controlled voltage source $V_{CH}$. Moreover, four capacitors $C_{CH,D}$, $C_{CH,S}$, $C_{G,CH}$ and $C_{SUB,CH}$ are inserted in the model to electrostically couple the *GNR* channel to Gate, Source, Drain and Substrate, respectively. Oxide thickness of *GNRFET* ($T_{OX}$) is inversely to $C_{G,CH}$ because a smaller $T_{OX}$ implies a larger oxide capacitance ($C_{OX}$), which yields more efficient controller of the channel potential. Therefore, $I_{ON}$ is increased and $I_{OFF}$ is reduced. The drain–source current $I_{DS}$ *N-GNRFET* is given by

$$I_{DS}(\psi_{CH}, V_D, V_S) = \frac{2qkT}{h} \sum_{\alpha} \left[ ln\left(1 + e^{\frac{q(\psi_{CH}-V_S)-\varepsilon_\alpha}{KT}}\right) - ln\left(1 + e^{\frac{q(\psi_{CH}-V_D)-\varepsilon_\alpha}{KT}}\right)\right], \quad \rightarrow \quad (1)$$

where $\Psi_{CH}$ is channel potential and it is determined by the electrostatics in the channel, $\varepsilon_\alpha$ is density of electrons in a given subband, $q$ is the electron charge, $K$ is the Boltzmann's constant, $h$ is the reduced Planck's constant, and $T$ is the temperature. $Q_{CH}$ is channel charge and $Q_{CAP}$ is charge across the different capacitors that couple into the channel. $\Psi_{CH}$ is the negative of the intrinsic energy level ($E_i$) and thus the conduction band $E_C = \varepsilon_\alpha - \Psi_{CH}$ and the valence band $E_V = -\varepsilon_\alpha - \Psi_{CH}$.

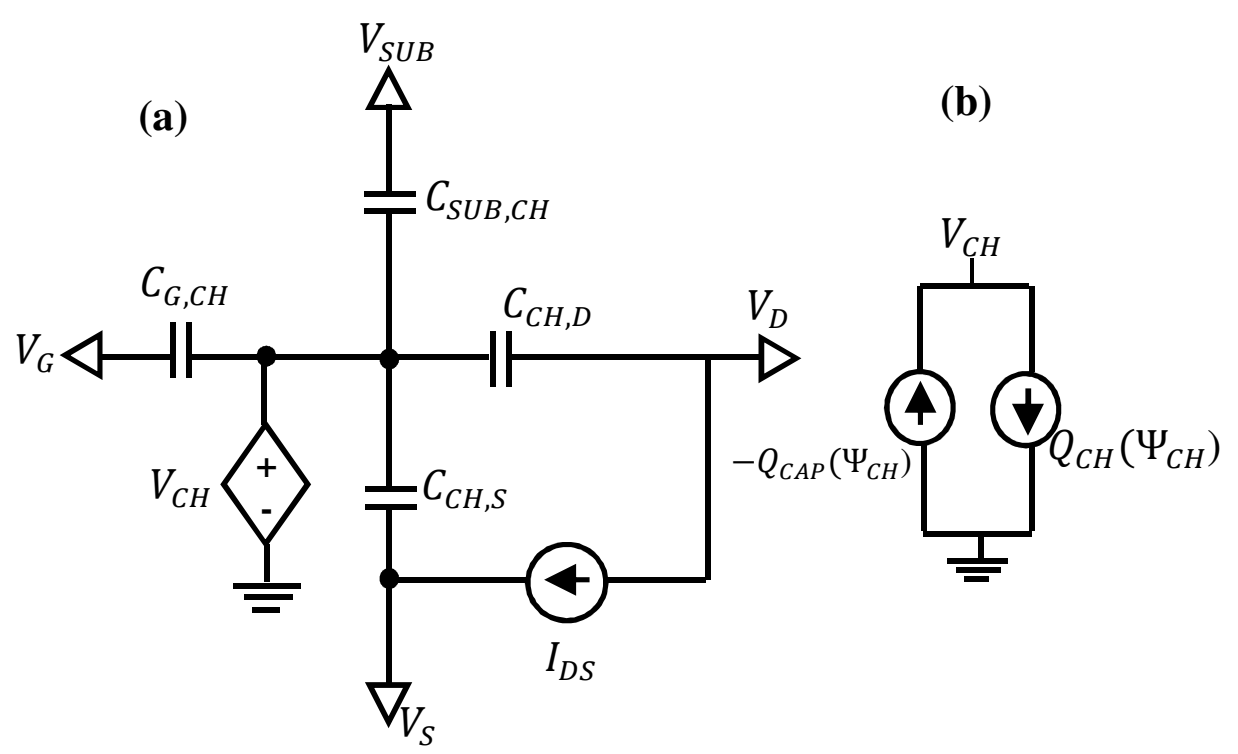

Figure 3. (a) Equivalent circuit model of *GNRFET*, and (b) Equation Solver.

In this paper, *GNRFET* is employed as a power switch instead of *MOSFET* in different power gating technique as presented in the following sections.

# III. Power Gating Based on GNRFET/MOSFET Conjunction

Power gating has been introduced to reduce sub-threshold leakage as well as gate leakage[16]. Scaling *MOSFET* to smaller physical dimensions causes scaling down for some parameters such as threshold voltage and oxide thickness which leading to increase the sub-threshold leakage current exponentially. *MOSFET* transistor with a *low*-$V_{th}$ and a thin oxide suffers from the gate leakage much more seriously than a transistor with a *high*-$V_{th}$ with a thick oxide as a result of an exponential dependence on the oxide thickness. Hence, the traditional power gating structures tend to use *low*-$V_{th}$ MOSFET transistors in the power gating modules to keep fast performance in active mode and *high*-$V_{th}$ MOSFET as power switch to reduce the leakage current in sleep mode in the advanced technologies. The main problem of *MOSPS* in active mode (*high*-$V_{th}$ power switch turns ON), the Virtual Ground rail ($VGNR$) becomes less than the supply voltage due to the IR drop across the power switch which leads to increasing in logic delay. While in Sleep mode (power switch turns OFF), the sub-threshold leakage current decreases exponentially in with increased $V_{th}$. This problem may be not a big issue in the combinatorial circuits but it cause to lost all data stored in the flip-flop for sequential circuits. Therefore, the extra circuits are used for the sequential circuits to store state values during sleep mode [17]. The proposed power gating structures based on *GMC* address the low-performance issue caused by *high*-$V_{th}$ of the conventional *MOSPG* structure at the low voltage while keeping low leakage-power dissipation in sleep mode. The structure of the *GMCPG* with single power switch and network power switch is discussed in the following subsections *A* and *B*, respectively.

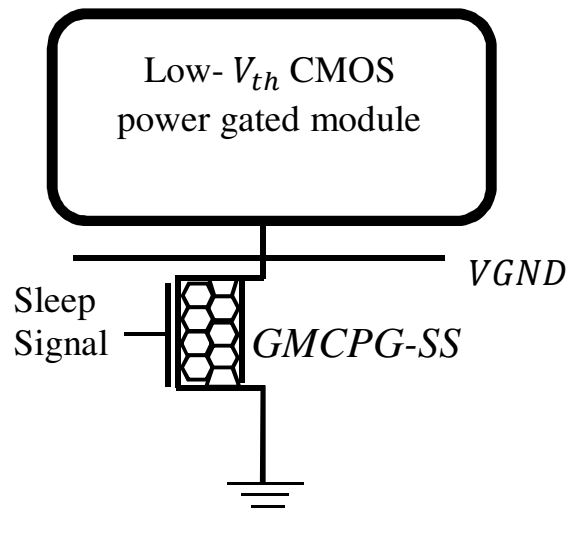

Figure 4. *GMCPG-SS structure*

### *A. GMCPG with GNR Single Power Switch (GMCPG-SS)*

As shown in Figure 4, the structure of *GMCPG-SS* depends on connect the *low*-$V_{th}$ *CMOS* power gated module series with *N-GNRFET* sleep transistor as a footer with connecting the Substrate terminal of the *N-GNRFET* to a ground. Depending on $V_{GS}$, a *GNRPS* operates either in the Active mode or in Sleep mode. The power gated module operates regularly when *GNRPS* turns ON. When *GNRPS* is turned OFF, the voltage level of $VGNR$ is varied to a value near $V_{cc}$ to suppress the leakage power of the power gated module. When *GNRPS* makes the transition again from Sleep to Active mode, the parasitic capacitance at $VGNR$ node should be completely discharge through *GNRPS* to return the voltage level of $VGNR$ to the nominal value. This process needs a relatively long wake-up time but it still less than that of *MOSPG* as presented in Section V. It is undoubted that the structure of *GMCPG-SS* is similar to the structure of traditional power gating but the main differences reside into the parameters of *GNRPS*. The model parameters of *GNRFET* are channel length ($L_{CH}$), the reservoir length ($L_{RES}$), the ribbon width ($W_{CH}$), the gate width ($W_G$), and the spacing between the ribbons ($W_{SP}$). *GNRFET* is characterized by tanning bandgap which there is inversely proportion between the induced energy bandgap and the width of *GNR*. Hence, increasing the bandgap means decreasing the width of Graphene, while the bandgap of Silicon *MOSFET* is invariable and is not depend on the thickness of oxide layer. The width of a *GNR* is given by

$$W_{CH} = (N+1)\sqrt{3} \times \frac{dcc}{2}, \tag{2}$$

where $N$ is dimer lines and $dcc$ is carbon to carbon bond distance. The gate width $W_G$ is determined based on $W_{CH}$ as following

$$W_G = (2W_{SP} + W_{CH}) \times n_{rip}, \tag{3}$$

where $W_{sp}$ is the spacing between ribbons and $n_{Rib}$ is the number of ribbons. The impact of *GNRFET* device parameters (the number of dimer lines, the number of ribbon, and the spacing between ribbons) on the circuit delay and the leakage power for *GMCPG-SS* is evaluated and presented in Section V. Where *GMCPG-SS* contains from *N-GNR* power switch as a footer and the power gated module. The power gated module consists of 20 *low*-$V_{th}$ MOSFET inverter chains and each chain consists of 20 inverters. To minimize the wake-up time and reduce the leakage power in *PG* structure, the network power switch is implemented in next subsection.

### B. *GMCPG* with *GNR Network Power Switch (GMCPG-NS)*

Significant difficulties of power switch with large width leads to a meaningful area overhead. Consequently, the power consumption is increased because of the considerable amount of leakage current in Sleep mode [18]. Several solution are presented using *MOS* power switch such as [19].To mitigate the aforesaid problem, *GNR* network power switch is presented and demonstrated in Figure 6. *GMCPG-NS* consists of three *N-GNRFET* transistors ($ST_1$, $ST_2$, and $ST_3$) and one *P-GNRFET* transistors ($ST_4$). $ST_1$ and $ST_2$ are the key sleep transistors while $ST_3$ and $ST_4$ are the control transistors with small sized. The network power switch is connected in series with *low-*$V_{th}$ *CMOS* power gated module as shown in Figure 6. During Active mode, $ST_3$ is turned OFF, and $ST_4$ is turned ON. Therefore, the sleep signal is asserted $ST_2$ for suppressing the gate tunneling leakage current of *GMCPG-NS* structure by the effect of low drain to source voltage ($V_{DS}$). Because of decreasing the drain voltage leads to an exponential decrease of the tunneling leakage current [5]. During Sleep mode, Gate terminal of $ST_2$ is connected to $ST_3$ by $n_1$ terminal, with taking under consideration that $V_{GS}$ of $ST_2$ is approximately equals "0", and $ST_2$ is turned OFF. Consequently, the leakage current of transistor $ST_1$ decreases proportionally with reducing the voltage difference between $VGNR$ and Gate terminal of $ST_2$ from the power supply voltage ($V_{CC}$) to $V_{CC} - n_1$. There is averting of gate leakage by $ST_1$ despite of the continuous flowing of the reduced gate leakage through $ST_3$ to the terminal $n_2$ as a result of $I_{g2}$ is less than that of *GMCPG-SS*. The simulation results of *GMCPG-NS* and compared with *GMCPG-SS* are presented in Section V. To improve the wake-up time of *GNR* power switch, the multimode *PG* structures using *GNRPS* are provided in next Section.

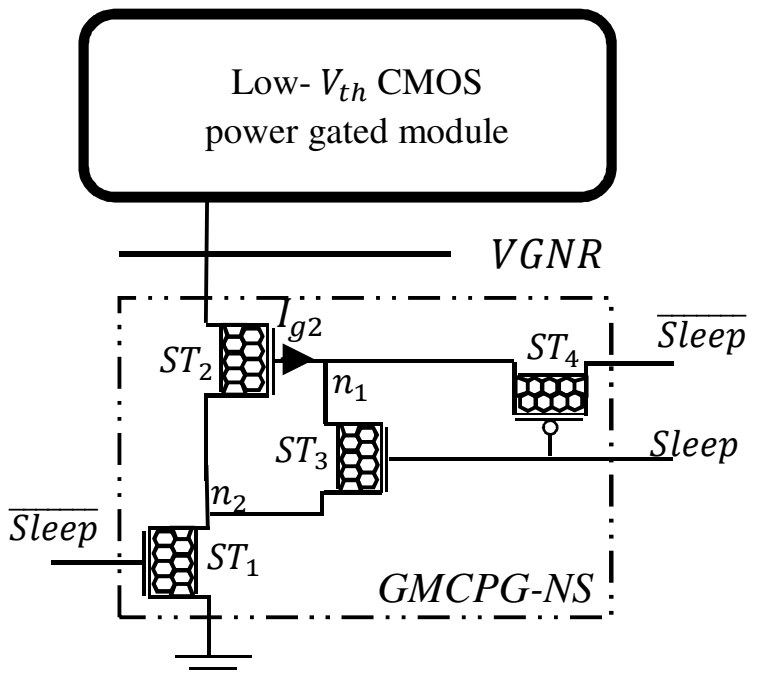

Figure 5. *GMCPG-NS structure*

## IV. MULTIMODE *GMCPG* STRUCTURES

PG structures struggle with the large wake-up time during the transitioning from Sleep to Active modes, particularly with numerous power mode transitions over short intervals resulting in greater wake-up power. Therefore, the required solution is using power gated modules into multiple power saving modes which compensates small increase of leakage power dissipation by a reduction in wake-up time. Two different multimode structures based on *GMC* are built and compared with the same structure based on *MOS* power switch. In Subsection IV.A. Triple Modes *GMCPG* structure is demonstrated. Quadratic modes *GMCPG* structure is provided in Subsection IV.B.

### A. *Triple- Modes GMCPG*

Triple- Modes GMCPG (*TM-GMCPG*) structure is shown in Figure 7. The power switching of *TM-GMCPG* consists of *P-GNRFET* ($ST_P$) in parallel to *N-GNRFET* ($ST_N$). The control signals are NSS and PSS as shown in Figure 7. The proposed *TM-GMCPG* presents three different power modes; Active, Nap and Sleep. During transition of power-modes, fluctuations of leakage power and power supply magnitude are reduced by the intermediate mode. The various modes of operation are listed in Table 1.

According to the control signals values inserted in Table 1, $VGNR$ equals Ground supply voltage ($V_{SS}$) which leads to high-speed operation in Active mode of *TM-GMCPG*. During Nap mode, the logic "0" is assured for each control signal as reported in Table 1, which leads to turn OFF $ST_N$ and force $ST_P$ to operate as a source-follower. Consequently, $VGNR$ equals the gate source voltage ($V_{GSP}$) of $ST_P$ which means that the applied voltage on *CMOS* logic module becomes $V_{CC} - |V_{GSP}|$. The estimated value of applied voltage on *CMOS* logic module reduces the leakage currents as a result of the proportional dependent between both factors. During sleep mode, both sleep transistors are turned OFF according to the values of control signals as listed in Table 1. As a result, $VGNR$ node is close to $V_{CC}$ which means that the path to *GND* is disconnected. Increasing the $V_{GS}$ and reducing leakage currents are the considerable outcome from *DIBL* [20- 21].The leakage current and wake-up time for *TM-GMCPG* under different *CMOS* logic modules are presented in Section *V*. Another multimode *PG* based on *GNRPS* is presented in next subsection.

**Table 1. The different modes for *TM-GMCPG***

| | $\overline{\text{NSS}}$ | $\overline{\text{PSS}}$ | $ST_N$ | $ST_P$ |
|---|---|---|---|---|
| Active mode | 1 | 1 | ON | OFF |
| Nap mode | 0 | 0 | OFF | ON |
| Sleep mode | 0 | 1 | OFF | OFF |

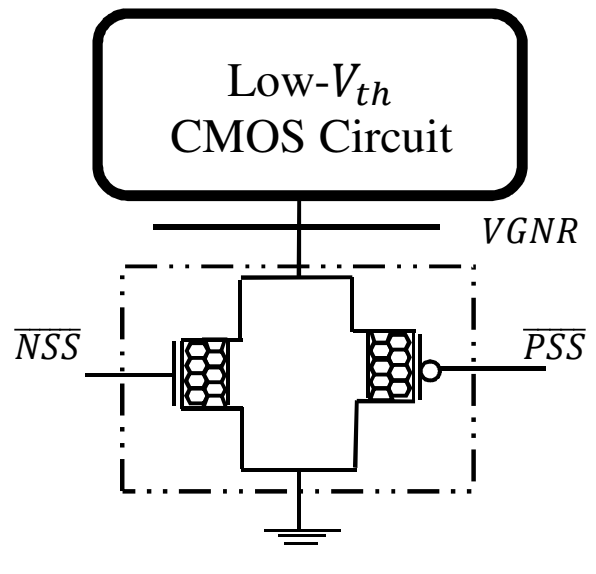

Figure 6. *TM-GMCPG* structure

### B. *Quadratic Modes GMCPG*

The proposed Quadratic Modes *GMCPG* (*QM-GMCPG*) structure presents four different power modes. The power modes are Active, Nap, Slumber and Sleep. The *QM-GMCPG* is shown in Figure 7. The power switching of *QM-GMCPG* structure consists of the main *GNRPS* $ST_0$ and two small *GNRPSs* $ST_1$ and $ST_2$, each corresponding to intermediate modes (Nap and Slumber). The control signals are $\overline{\text{SPS}}$, $\overline{\text{NPS}}$ and SLS as shown in Figure 8. The control signals are the key parameters to operate the certain mode. The various modes of operation are listed in Table 2. In Active mode, logic "0" is assured to control signals $\overline{\text{NAP}}$ and $\overline{\text{SLS}}$ which mean that $V_{BG}$ of each transistor is "0". In addition, logic "1" is set to control signal $\overline{\text{SPS}}$ so that $ST_0$, $ST_1$ and $ST_2$ transistors are ON which leads to high-speed operation and increasing the leakage power compared to the other modes. In Nap mode, SLS and $\overline{\text{SPS}}$ are set "1" which leads to make transistors $ST_0$ and $ST_2$ are OFF and $ST_1$ is partially ON by forward back-gate biasing. During Slumber mode, transistor $ST_2$ is partially ON as a result of set logic "0" to control signals $\overline{\text{NPS}}$ and $\overline{\text{SPS}}$ as reported in Table 2. With taking under consideration that the number of dimer lines is the crucial factor for making a difference between $ST_1$ and $ST_2$ in the passing current. Consequently, the current passing through $ST_2$ is less than that of $ST_1$ since $N_1 > N_2$ so that $VGNR$ in Nap mode is lower than that in Slumber mode. This technique is very important to switch between both intermediate modes. All transistors are turned OFF during Sleep mode when all control signals are set to "0". During Sleep mode, there is a small leakage current is flowing through the transistor and wake-up time is increased as compared to the other power saving modes. The leakage power and wake-up time for *QM-GMCPG* under different *MOS* power gated modules are presented in Section *V*.

Table 2. The different modes for *QM-GMCPG*

| | NPS | SLS | SPS | Transistors Status |
|---|---|---|---|---|
| Active mode | 0 | 0 | 1 | All *GNRPS*s ON |
| Nap mode | 1 | 0 | 0 | $ST_1$ partially ON |
| Slumber mode | 0 | 1 | 0 | $ST_2$ partially ON |
| Sleep mode | 0 | 0 | 0 | All *GNRPS*s OFF |

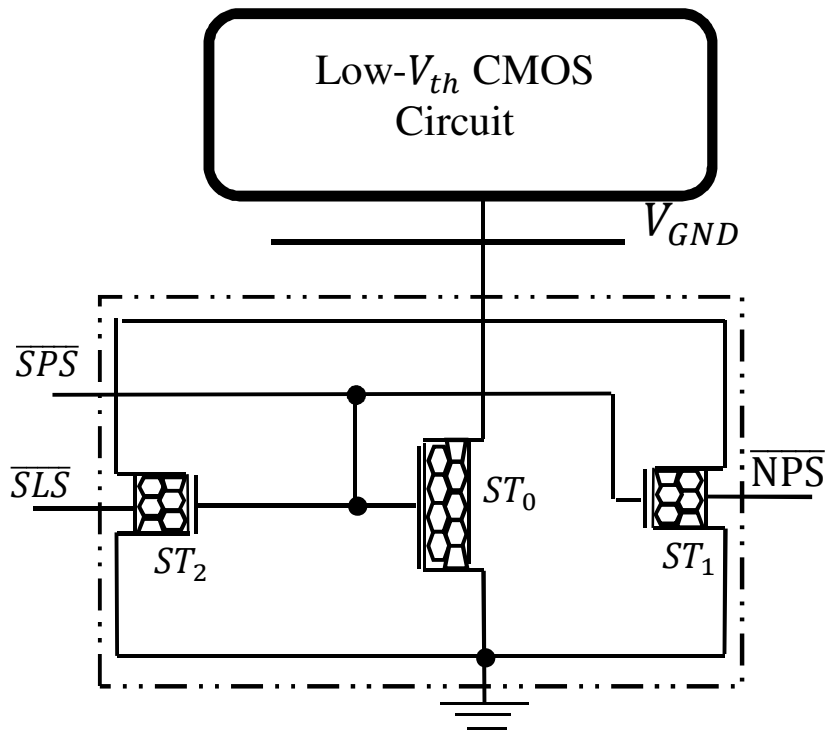

Figure 7. *MQ-GMCPG* structure

## V. SIMULATION RESULTS

HSPICE tool is used to build *GMCPG* structures. Moreover, HSPICE is used to implement ISCAS85 benchmark circuits as *CMOS* power gated modules in *GMCPG* structure using fabrication technology 16 nm predictive technology (PTM) model [22]. The supply voltage is 0.7 V, 0.35V for *MOSFET* and *GNRFET*, respectively. ISCAS85 circuits are implemented using *MOS* transistors with a high-k metal gate strained-si with low threshold voltage $V_{th} = 0.47965V$ and $t_{ox} = 0.95nm$ for *NMOS*, and $V_{th} = -0.43121V$ and $t_{ox} = 1nm$ for *PMOS* [23]. *GNRPS*s are implemented using a 16nm NanoTCAD ViDES HSPICE model [23]. Performance of proposed *GMCPG* design is investigated in terms of leakage power, delay and wake-up time.

The characteristics of *GNRFET* transistor is demonstrated in Subsection *A*. In Subsection *B*, the performance parameters are evaluated for *GMCPG-SS* and compared with the convention *MOSPG* under different power gated modules. The circuit delay and the leakage current using *GMCPG-NS* for different ISCAS85 benchmark circuits are presented in Subsection *C*. In Subsection *D*, the performance parameters for different *CMOS* power gated modules using *TM-GMCPG* and *QM-GMCPG* are presented. In addition, power delay producer both multimode designs is calculated and presented in Subsection *D*.

### A. Characteristics of GNRFET

The drain current characteristics of a *N-GNRFET* and *N-MOSFET* are demonstrated as a function of drain-to-source voltage ($V_{DS}$) and gate to source voltage ($V_{GS}$) as shown in Figure 4 (a) and (b), respectively. The parameters of GNRFET are reported in Table 3. Using Equations 2 and 3, the width of *GNRFET* is $33.6nm$. Therefore, the 16-nm *N-GNRFET* with gate width $33.6nm$ is compared to 16-nm N-type *MOSFET* with same gate width for fair comparison.

Gate voltage ($V_g$) is exploited to manage *GNRFET*. As shown in Figure 8 (a), the drain current ($I_{DS}$) of the *GNRFET* is exceeding that of *MOSFET*. In addition, in the low electron concentration, the Drain-Induced Barrier Lowering (*DIBL*) and Gate-Induced Drain Leakage (*GIDL*) effects are approximately minuscule. As shown in the Figure 8, *GNRFET* outperforms *MOSFET* transistor because the former has a larger ON current and lower leakage current than *MOSFET* transistor. With increasing $V_{GS}$, $I_{DS}$ of the *GNRFET* is saturated at higher $V_{DS}$ as compared to *MOSFET*. Consequently, $V_{th}$ of *GNRFET* is lower than *MOSFET* by $45\%$ as shown in Figure8 (b), where $I_{DS}$ of the *GNRFET* is greater than that of the *MOSFET*. Moreover, the switch efficiency ratio of *GNRFET* is higher up to 44.1 as compared to *MOSFET*.

Table 3. Required parameters of GNRFET

| GNRFET parameters | Value |
|---|---|
| Source/drain region Length | 16nm |
| Channel length | 16nm |
| Number of GNR | 6 |
| Number of dimer lines | 12 |
| Oxide Thickness of top gate | 0.95nm |
| Oxide Thickness between channel and substrate | 20nm |
| Doping Fraction | 0.001 |
| Spacing between the edges of two adjacent GNRs | 2nm |

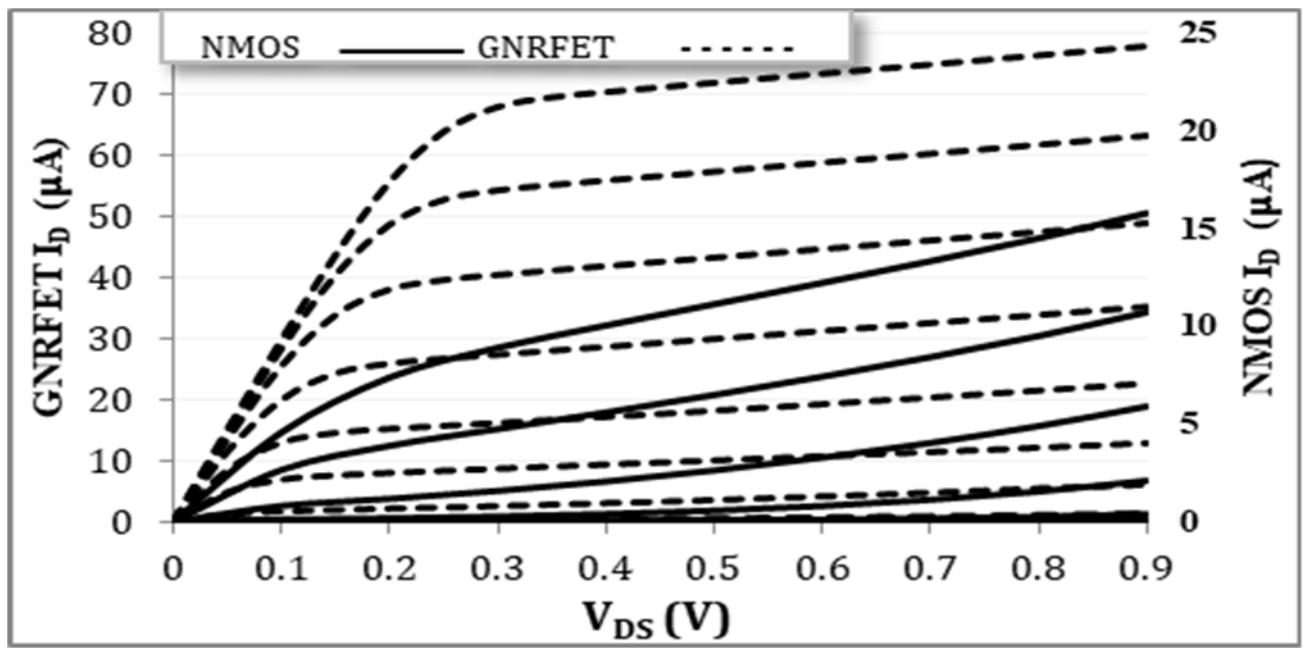

(a)

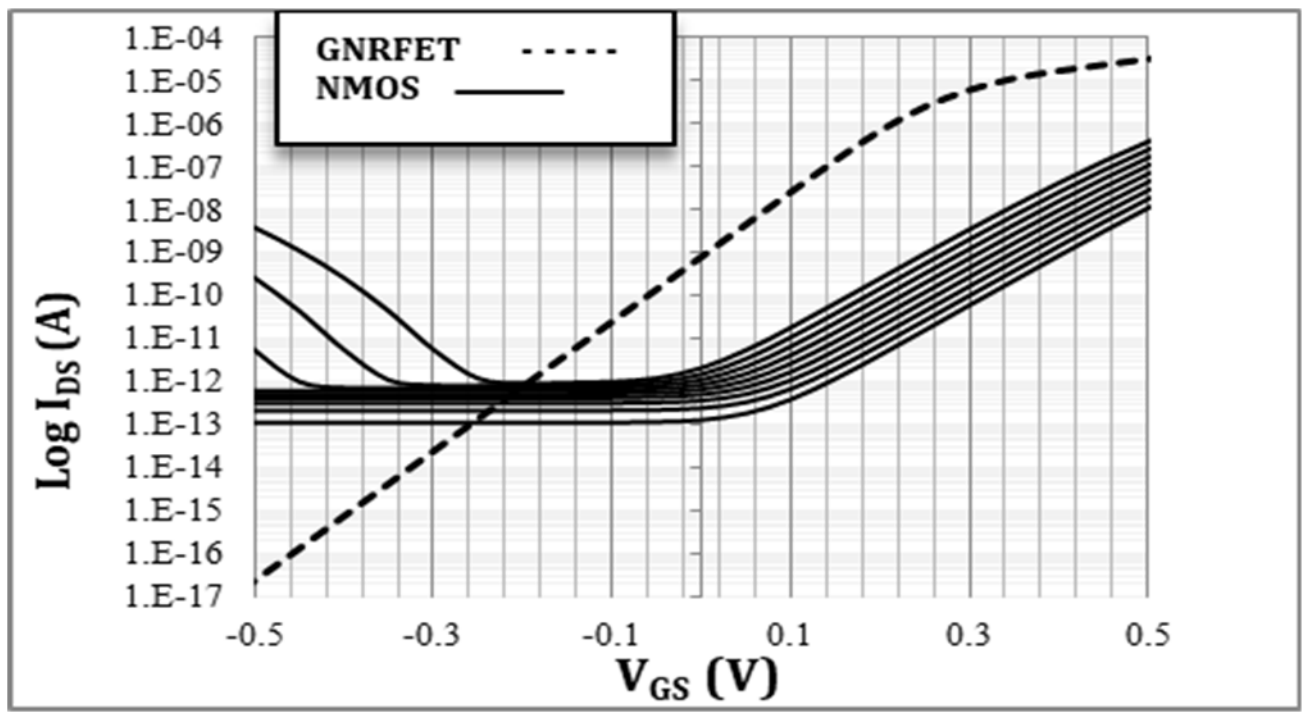

(b)

Figure 8. Drain current of a $16nm$ *N-GNRFET* and a $16nm$ *N-MOSFET* as a function of (a) drain-to-source voltage for different gate-to-drain voltage, (b) gate-to-source voltage for different drain-to-source voltage, where *GNRFET* parameters are $N = 12, n_{Rib}= 6$ and $t_{ox} = 0.95nm$ and *NMOS* Parameters are $W = 33.6\ nm$ and back-gate voltage =0

From the other hand, the impact of *GNRFET* device on the circuit delay and the leakage power for *GMCPG* structure is calculated. The power gated module consists of 20 *low-*$V_{th}$ MOSFET inverter chains and each chain consists of 20 inverters. The power gated module connects to *N-GNR* power switch. The inverters chains are designed using 16 nm MOS transistors. The power switching size is designed as 10% of the total *NMOS* width in the inverter chains to evaluate the delay and leakage power for the power gated module. Using Equations (2) and (3), the number of dimer lines, the number of *GNR*s, and the spacing between *GNR*s are 75, 12 and 4 nm, respectively. The impact of *GNRFET* device on the circuit delay and the leakage power is shown in Figure 9. As shown in Figure 9 (a), the delay and leakage power are affected by changing in the number of dimer lines. The leakage power is exponentially increased at *N* up to 12. Delay is approximately constant with the changing number of *GNR* below 140 with rapidly increase in leakage power as shown in Figure 9 (b). The simulation results illustrate that the delay and leakage power are approximately not affected by the changing in the spacing between ribbons from 4nm to 10nm as shown in Figure 9 (c). Based on the simulation results, the determination of the optimum values for the model parameters of *GNRPG* leading to increase the switching speed in Active mode and reduce the leakage current in Sleep mode. The delay, leakage power and wake-up time for *GMCPG-SS* are evaluated for different power gated modules and reported in next subsection.

### B. Performance Parameters for GMCPG-SS

Performance parameters of *GMCPG* with using *GNR* single power switching using different power gated modules (ISCAS85 Circuits) are listed in Table 4 [24]. The percentages of variability for the performance parameters using proposed *GMCPG-SS* compared with nominal values using convention *MOSPG* are evaluated and reported in Table 4. The size ratio of *PS* width as compared to total *NMOS* width of each power gated module is 1% and 10% for *GNRPS* and *MOSPS*, respectively. Therefore, *GMCPG-SS* outperforms the conventional *MOSPG* regarding the area prospective which can be 95%. As shown in Table 3, due to *GMCPG-SS*, the leakage power of different power gated modules is decreased with at least 69% relative to the *MOSPG*. *GMCPG-SS* saves both delay time and wake-up time at least by 14% for each one as compared to the conventional *MOSPG* structure. On the average, *GMCPG-SS* reduces the leakage power, delay time and wake-up time on average up to 88%, 44%, and 24%, respectively, for different ISCAS85 power gated modules as compared to the *MOSPG*. Reducing the wake-up time by 14% as compared to *MOSPG* means increasing the switching efficiency by 44.1% compared to *MOSPG*. In addition, Power Delay Product (PDP) for each power gated module is calculated and demonstrated in Figure 10. *GMCPG-SS* minimizes PDP value as compared to conventional *MOSPG* between 84% and 94% as shown in Figure 10. To reduce the leakage current another design for *GNR* power switching is investigated and provided in the next subsection.

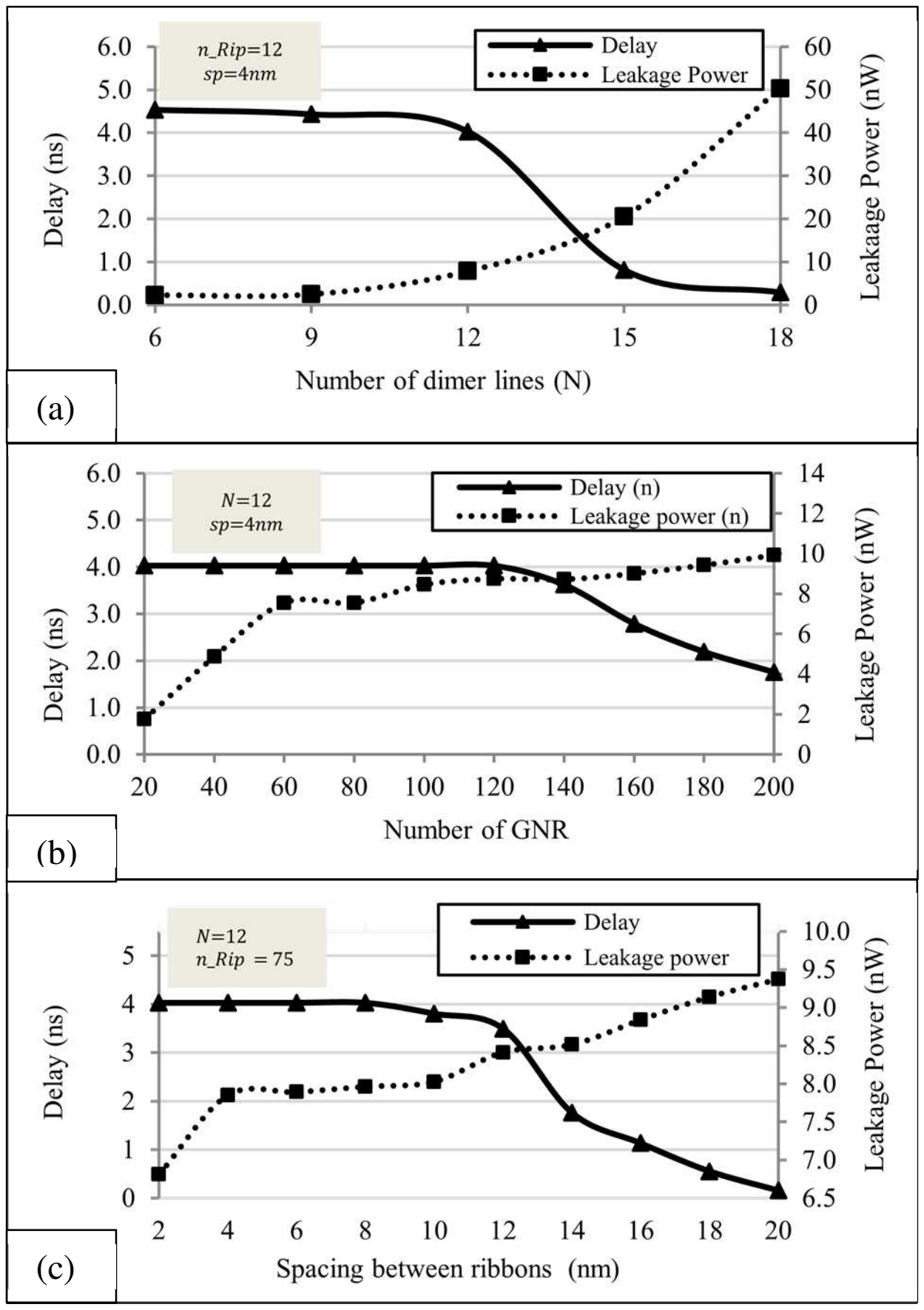

Figure 9. Delay and leakage power as a result of changing *GNRFET* parameters (a) The number of dimer lines, (b) The number of *GNR*s, and (c) The spacing between *GNR*s.

*C. Performance Parameters for GMCPG-NS*

*GNR* network power switch is essential to reduce the leakage current and wake-up time as compared to *MOSPG* and *GMCPG-SS*. To demonstrate the effectiveness of the proposed *GMCPG-NS*, the leakage power, delay and wake-up time of the *GMCPG-NF* under different power gated modules are evaluated as compared to *MOSPG* and demonstrated in Table 5 [24]. On the average, *GMCPG-NS* reduces the leakage power, delay and wake-up time by 86%, 45% and 37% in different power gated modules ISCAS85 as compared to *MOSPG* structure. Although *GMCPG-NS* uses four transistors in the power switch structure, the design still has the lower area up to 87% as compared to *MOSPG*. In addition, the performance of *GMCPG-NS* is compared to *GMCPG-SS* to evaluate the improvement as a result of using the network power switching. On the average, *GMCPG-NS* is reduced the leakage power, delay and wake-up time by 30%, 3% and 15% respectively, as compared to *GMCPG-SS*. On the average, *GMCPG-NS* achieves PDP reduction up to 90% and 54% as compared to *MOSPG* and *GMCPG-SS*, respectively as shown in Figure 10. The performance parameters are improved by using *GMCPG* multi-modes as presented in next subsections.

*D. Performance Parameters for Multimode GMCPG*

Leakage power and wake-up time are used to evaluate the performance of *TM-GMCPG* and *QM-GMCPG* under various power gated modules. The leakage power and wake-up time for *TM-GMCPG* as compared to the other *PG* structures are demonstrated on Figure 11 and 12, respectively. For fair comparison between various *PG* structures, all the demonstrated results in the both Figures are calculated in Sleep mode. *TM-GMCPG* outperforms the conventional *MOSPG* from the leakage power prospective. Consequently, the leakage power is reduced up to 87%, 37% and 19% as compared to *MOSPG*, *GMCPG-SS*, and *GMCPG-NS* respectively. On the wake-up time prospective, *TM-GMCPG* still able to reduce the wake-up time by 99% as compared to the others power gating structures as shown in Figure 12. On the average, during different power saving modes, the leakage powers are $2.12\mu W, 1.4\mu W$ and wake-up times are $3.77ns$, $4.22ns$ in Nap and Sleep modes, respectively. Therefore, the power saving mode is selected based on the requirement of the application of power gated module. Therefore, the number of power saving modes reflects the performance of multi-modes power gating in terms of leakage power and wake-up time. The leakage power of *QM-GMCPG* is improved on the average up to 85% versus *GMCPG* and 25% as compared to *GMCPG-SS* and by 2% as compared to *GMCPG-NS*. Due to different operation modes, for *QM-GMCPG* reduces the wake-up time approximately 99% as compared to the others *PG* structures. From the other point of view, the performance of *QM-GMCPG* is compared *TM-GMCPG.* The leakage power of *QM-GMCPG* is increased on average up to 44% while the wake-up time is reduced up to 24% as compared to *QM-GMCPG*. Therefore, increasing number of power saving modes leads to degrade the wake-up time with the expenses of leakage power.

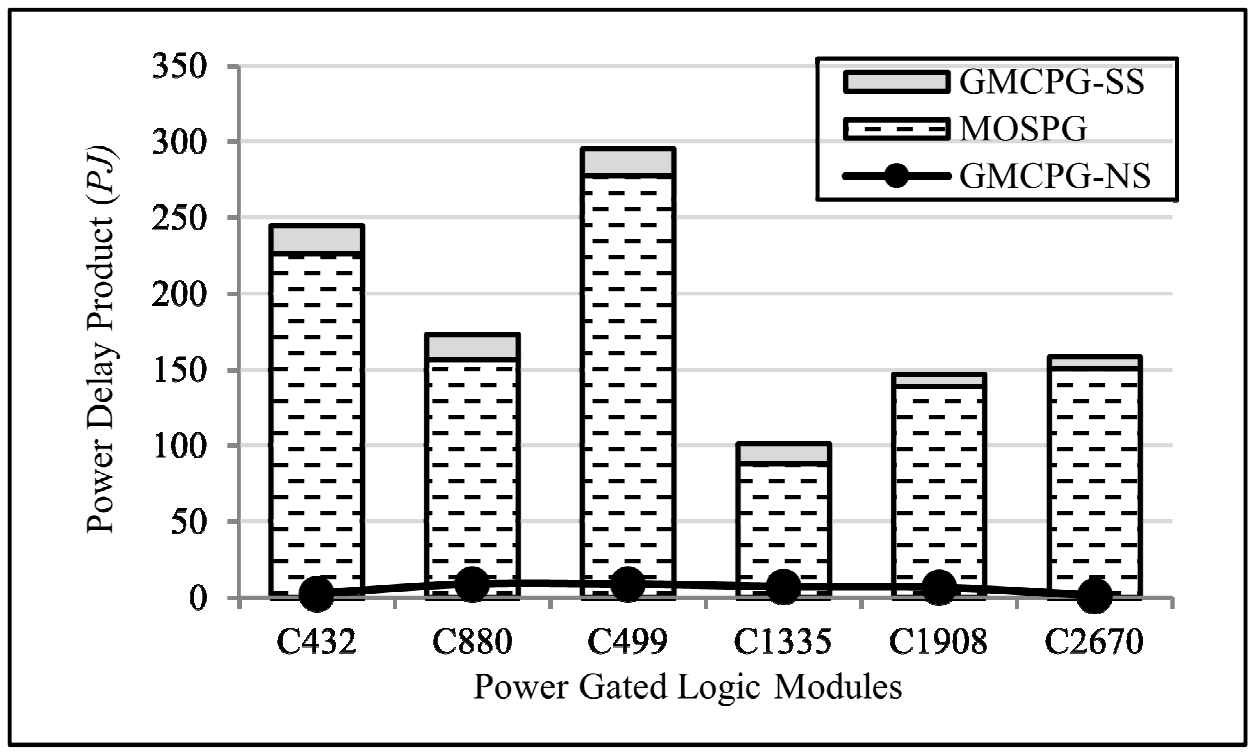

Figure 10 . Power Delay Product ($PJ$)

**Table 4.** Performance parameters of *GMCPG -SS* using different ISCAS85 Circuits at ($V_{CC} = 0.35$)

| Logic circuit | Leakage power (w) | | | Delay time (s) | | | Wake-up time (s) | | |
|---|---|---|---|---|---|---|---|---|---|
| | *MOSPG* | *GMCPG-SS* | Normalized value | *MOSPG* | *GMCPG-SS* | Normalized value | *MOSPG* | *GMCPG-SS* | Normalized value |
| **C432** | 5.39E-06 | 1.66E-06 | 69.2% | 6.31E-06 | 5.47E-06 | 13.34% | 4.00E-06 | 2.97E-06 | 25.82% |
| **C880** | 1.41E-05 | 1.92E-06 | 86.4% | 5.01E-06 | 3.19E-06 | 36.47% | 2.51E-06 | 1.82E-06 | 27.49% |
| **C499** | 2.75E-05 | 3.68E-06 | 86.6% | 1.07E-05 | 3.25E-06 | 69.77% | 2.50E-06 | 1.88E-06 | 24.80% |
| **C1335** | 2.38E-05 | 4.09E-06 | 82.8% | 3.33E-06 | 2.45E-06 | 26.31% | 4.05E-06 | 3.08E-06 | 23.95% |
| **C1908** | 1.77E-05 | 2.41E-06 | 86.4% | 5.13E-06 | 2.33E-06 | 54.46% | 7.62E-06 | 5.51E-06 | 27.73% |
| **C2670** | 1.97E-05 | 3.85E-06 | 80.5% | 5.10E-06 | 1.94E-06 | 61.98% | 6.12E-06 | 5.28E-06 | 13.71% |
| **Average** | **88.4%** | | | **44%** | | | **23.92%** | | |

**Table 5.** Performance parameters of *GMCPG -NS* using different ISCAS85 Circuits at ($V_{CC} = 0.35$)

| Logic circuit | Leakage power (w) | | | Delay time (s) | | | Wake-up time (s) | | |
|---|---|---|---|---|---|---|---|---|---|
| | *MOSPG* | *GMCPG-NS* | Normalized value | *MOSPG* | *GMCPG-NS* | Normalized value | *MOSPG* | *GMCPG-NS* | Normalized value |
| **C432** | 5.39E-06 | 1.67E-06 | 69.02% | 6.31E-06 | 5.01E-06 | 20.60% | 4.00E-06 | 2.57E-06 | 35.91% |
| **C880** | 1.41E-05 | 1.82E-06 | 87.09% | 5.01E-06 | 3.62E-06 | 27.89% | 2.51E-06 | 1.34E-06 | 46.65% |
| **C499** | 2.75E-05 | 3.36E-06 | 87.78% | 1.07E-05 | 3.24E-06 | 69.80% | 2.50E-06 | 1.69E-06 | 32.28% |
| **C1335** | 2.38E-05 | 1.70E-06 | 92.86% | 3.33E-06 | 2.51E-06 | 24.56% | 4.05E-06 | 2.28E-06 | 43.78% |
| **C1908** | 1.77E-05 | 1.62E-06 | 90.85% | 5.13E-06 | 3.25E-06 | 36.60% | 7.62E-06 | 5.06E-06 | 33.62% |
| **C2670** | 1.97E-05 | 2.12E-06 | 89.24% | 5.10E-06 | 5.65E-07 | 88.93% | 6.12E-06 | 4.46E-06 | 27.11% |
| **Average** | **86.14%** | | | **45%** | | | **36.56%** | | |

PDP in Active mode for *TM-GMCPG* and *QM-GMCPG* is demonstrated in Figure 13 for each power gated module. As shown in the Figure 13, PDP of *QM-GMCPG* is lower than that of *TM-GMCPG* although the leakage power of the former is greater than *TM-GMCPG*. *QM-GMCPG* improves the PDP in between 5% and 72% as compared to *TM-GMCPG* under different power gated modules. From the other point of view, the multimode power gating structures reduce PDP in Active mode and the wake-up time as compared to the other PG structures.

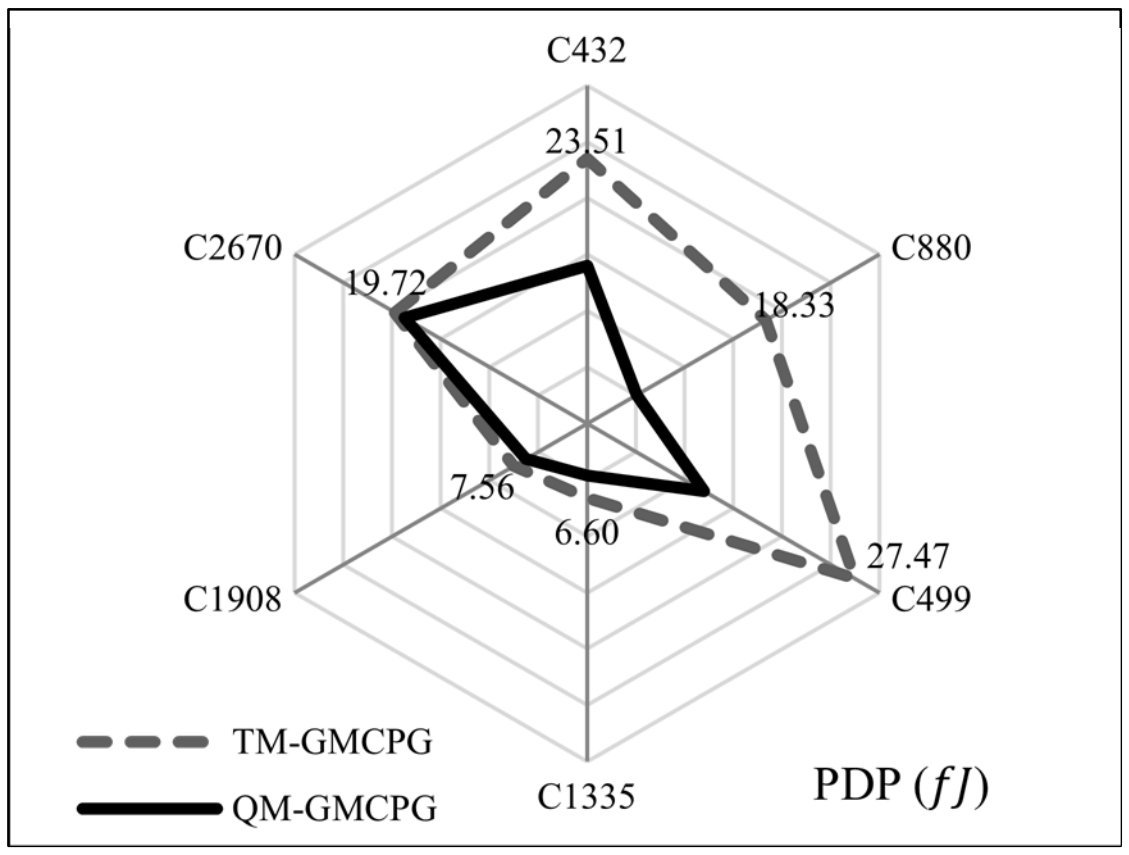

Figure 13. Power-Delay Product ($fJ$) in Active mode for *QM-GMCPG* and *TM-GMCPG*

## VI. CONCLUSION

A robust power gating structure is proposed based on *GNRFET/MOSFET* conjunction. Different PG structures based on *GMCPG* are considered. Various *CMOS* logic circuits are used as the power gated module in *GMCPG* structures. The proposed structures have several cons versus to conventional *MOS* power gating. The area overhead of *GMCPG* is smaller than *MOSPG*. Moreover, *GMCPG* structures are exploited to operate with *low*-$v_{th}$ *CMOS* circuits to address the problem of performance deterioration for *MOSPG* with advanced technologies. The performance circuit parameters of the proposed *GMCPG* structures outperform the conventional *MOSPG*. In addition, the wake-up time is reduced using *GMCPG* structures. *GMCPG-SS* structure reduces the leakage power, delay time and wake-up time on average up to 88%, 44%, and 24%, respectively, for different ISCAS85 power gated modules as compared to the *MOSPG* structure. On average, *GMCPG-NS* structure reduces the leakage power, delay and wake-up time by 86%, 45%, and 36% in different ISCAS85 power gated modules as compared to the *MOSPG* structure. On the average, *GMCPG-NS* achieves PDP reduction up to 90% and 54% as compared to *MOSPG* and *GMCPG-SS*, respectively. Two different multimode structures based on *GMC* are built to improve the wake-up time. The performance of *TM-GMCPG* and *QM-GMCPG* are evaluated using largest power gated modules. *TM-GMCPG* outperforms the *MOSPG* from the leakage power prospective. The leakage power of *TM-GMCPG* is reduced up to 87%, 37% and 18% as compared to *MOSPG GMCPG-SS* and *GMCPG-NS*, respectively. The leakage power of *QM-GMCPG* is improved on the average up to 85% versus *MOSPG,* 25% as compared to *GMCPG-SS* and 2% as compared to *GMCPG-NS*. While, *TM-GMCPG* and *QM-GMCPG* reduce the wake-up time approximately 99% as compared to the others *PG* structures. The leakage power of *QM-GMCPG* is increased on average up to44% while the wake-up time is reduced up to 42.7% as compared to *TM-GMCPG*. PDP of *QM-GMCPG* is lower than that of *TM-GMCPG* although the leakage power of the former is greater than *TM-GMCPG*. *QM-GMCPG* improves the PDP in between 5% and 72% as compared to *TM-GMCPG* under different power gated modules.

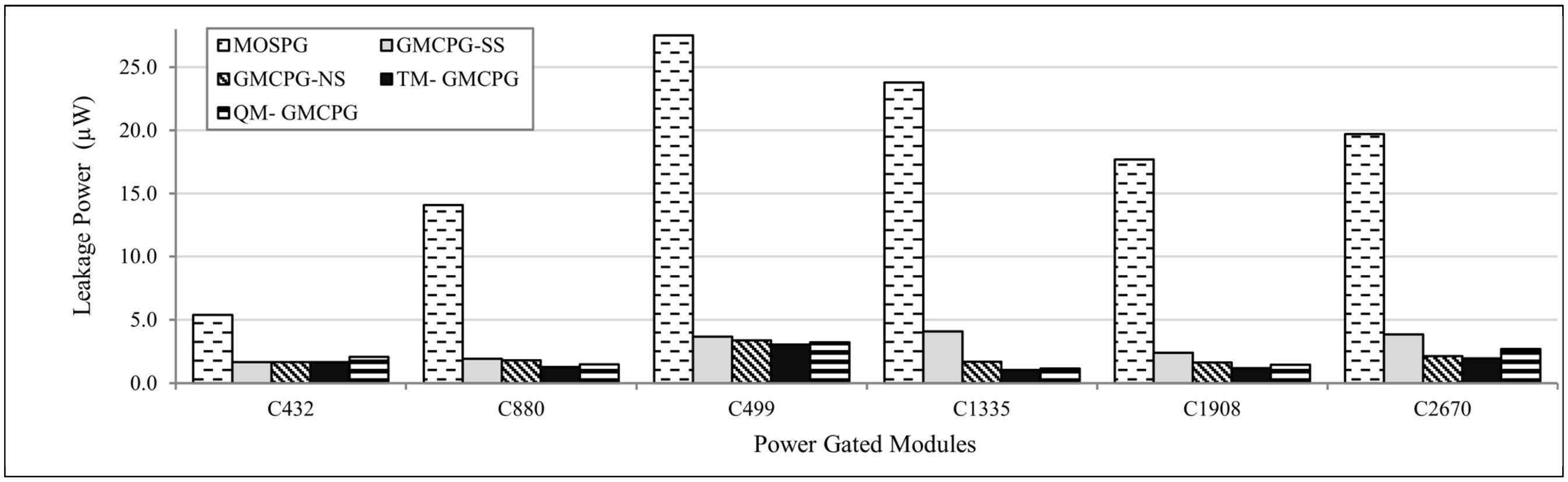

Figure 11. Leakage power for *TM-GMCPG* and *QM-GMCPG* as compared to conventional *MOSPG* and *GMCPG*

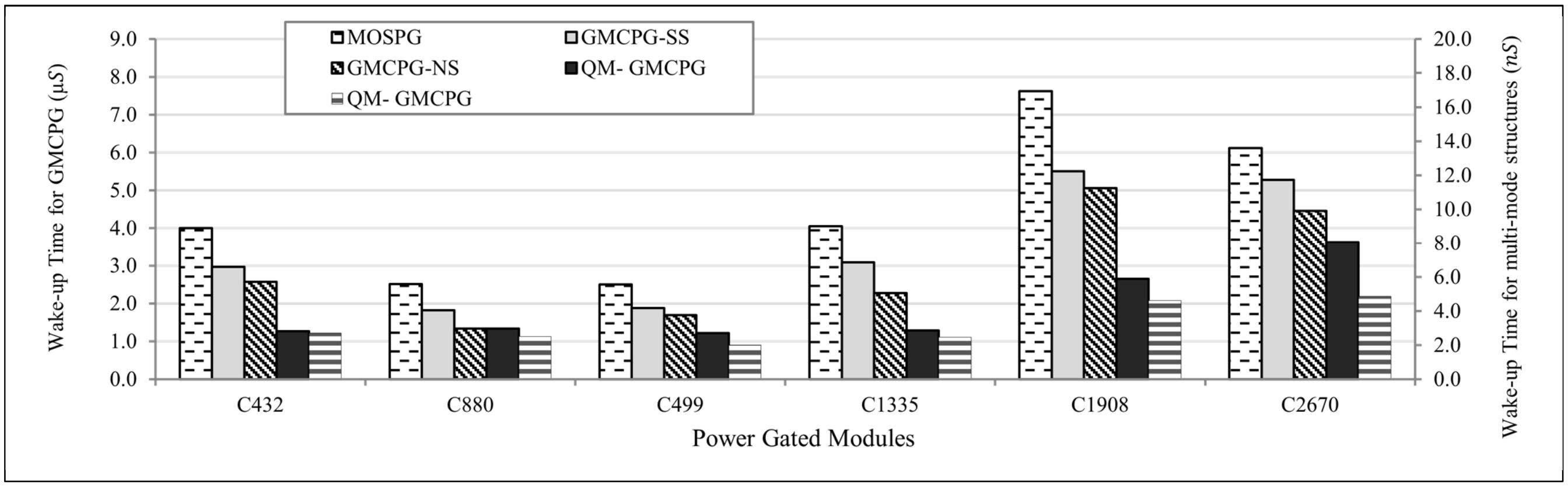

Figure 12. Wake-Up time for *TM-GMCPG* and *QM-GMCPG* as compared to conventional *MOSPG* and *GMCPG*